\documentclass[useAMS,usenatbib]{mn2e}
\usepackage{epsfig,subfigure,amsmath}
\usepackage{color}

\definecolor{purple}{rgb}{1,0,1}

\newcommand{\lcdm}{$\Lambda$CDM}
\newcommand{\hmpc}{$h^{-1}$Mpc}

\newcommand{\hmsol}{\mbox{ } {h}^{-1}~{M}_{\odot}}

\bibpunct[; ]{(}{)}{;}{a}{}{,}

\def\velociraptor{{\tt VELOCIraptor}}

\title{The life and death of cosmic voids}
\author[P.M. Sutter et al.]
{
\parbox{\textwidth}{
{P.M. Sutter}$^{1,2,3}$ \thanks{Email: sutter@iap.fr},
Pascal Elahi$^{4}$,
Bridget Falck$^{5}$,
Julian Onions$^{6}$,
Nico Hamaus$^{1,2}$,
Alexander Knebe$^{7}$,
Chaichalit Srisawat$^{8}$, and
Aurel Schneider$^{8}$
}
\vspace{0.4cm}\\
\parbox[c]{\textwidth}{
$^{1}$ Sorbonne Universit\'{e}s, UPMC Univ Paris 06, UMR7095, Institut d'Astrophysique de Paris, F-75014, Paris, France \\
$^{2}$ CNRS, UMR7095, Institut d'Astrophysique de Paris, F-75014, Paris, France \\
$^{3}$ Center for Cosmology and AstroParticle Physics, Ohio State University, Columbus, OH 43210\\
$^{4}$ Sydney Institute for Astronomy, University of Sydney, Sydney NSW 2016, Australia \\
$^{5}$ Institute of Cosmology and Gravitation, University of Portsmouth, Portsmouth PO1 3FX, UK  \\
$^{6}$ School of Physics \& Astronomy, University of Nottingham, Nottingham, NG7 2RD, UK  \\
$^{7}$ Departamento de F\'isica Te\'orica, M\'odulo C-15, Facultad de Ciencias, Universidad Aut\'onoma de Madrid, 28049 Cantoblanco, Madrid, Spain  \\
$^{8}$ Department of Physics \& Astronomy, University of Sussex, Brighton, BN1 9QH, UK
}}

\begin{document}

\maketitle

\label{firstpage}

\begin{abstract}
We investigate the formation, growth, merger history, movement, and 
destruction of cosmic voids detected via the watershed transform 
code \textsc{VIDE}
in a cosmological $N$-body dark matter 
\lcdm~simulation. By adapting a method used to construct halo merger trees, 
we are able to trace individual voids back to their initial appearance
 and record
the merging and evolution 
of their progenitors at high redshift. For the scales of void 
sizes captured in our simulation, we find that the void formation 
rate peaks at scale factor 0.3, which coincides with a growth in the 
void hierarchy and the emergence of dark energy. 
Voids of all sizes appear at all scale factors, though the median initial 
void size decreases with time. When voids become detectable they 
have nearly their present-day volumes.
Almost all voids have relatively stable growth rates and suffer 
only infrequent minor mergers.
Dissolution of a void via merging is very rare. Instead, most 
voids maintain their distinct identity as annexed subvoids of a larger 
parent.
The smallest voids are collapsing
at the present epoch, but void destruction ceases after 
scale factor 0.3.  
In addition, voids centers tend to move very little, 
less than 0.01 of their 
effective radii per $\ln a$, over their lifetimes. 
Overall, most voids exhibit little radical dynamical evolution;
their quiet lives make them pristine probes of cosmological 
initial conditions and the imprint of dark energy.
\end{abstract}


\section{Introduction}

Since cosmic voids are, by definition, relatively empty of matter, they 
offer a unique and pristine laboratory for studying 
dark energy~\citep{LavauxGuilhem2011,Sutter2012b},
exotic fifth forces~\citep{Li2012, Spolyar2013},
and the early universe~\citep{Goldberg2004}.
They also offer a complementary probe of the growth of structure via
their size and shape distributions~\citep{Biswas2010,Bos2012,Clampitt2013}.
Recently large catalogs
of voids identified
in galaxy redshift surveys~\citep{Pan2011,Sutter2012a,Nadathur2014,Sutter2013c}
have opened the way for
statistical and systematic measurements
of void properties~\citep{Ceccarelli2013,Sutter2013b}, 
and their
connections to cosmological parameters~\citep{Planck2013b, Melchior2013}.

However,
given the promising utility of voids, we still lack a detailed understanding
of their life cycles. For example, for a given void observed at low redshift,
we do not know when it formed, where it formed, whether it grew to its
present size via simple expansion or through mergers, nor whether it
will continue expanding or eventually collapse.
We also do not understand basic statistics about voids over
cosmic time: their formation and merger rates, growth rates,
and movement.
Such understanding of the life cycles of voids will solidify
current void-based cosmological analysis and enable future probes.
Also, if we are to use voids as cosmological probes we must understand
the impact of their dynamics on any primordial cosmological signal.

As identified theoretically by~\citet{Sheth2004} 
and discussed in the review of~\citet{VandeWey2011b}, void evolution appears 
intimately tied to its environment: smaller voids tend to appear inside 
larger overdense surroundings, while larger voids are truly 
anti-correlated with respect to the matter distribution. Thus smaller 
voids tend to collapse over time, while larger voids continue to 
expand. The expansion of larger voids causes their interiors to
 appear as miniature open universes, with lower-density walls, filaments, 
and halos~\citep{vdW2004, Aragon2010, Neyrinck2013} evolving in a 
self-similar pattern.
The evolutionary and hierarchical 
behavior has been modeled in the context of the adhesion 
approximation~\citep{Sahni1994}, simulations~\citep{vandeWey1993}, 
and excursion set theory~\citep{Sheth2004}.

However, void abundances are still difficult to predict with
excursion set formalisms alone~\citep{Jennings2013,Sutter2013a}.
While there have been several attempts to improve the initial
theoretical result of~\citet{Sheth2004}, such as by
adjusting the
void growth and destruction
parameters~\citep{Furlanetto2006,Daloisio2007,Paranjape2012} and
rescaling void sizes~\citep{Jennings2013}, there still
remains very little correspondence to voids identified with watershed techniques in galaxy surveys~\citep{Sutter2013c}.
We may improve excursion set predictions by directly measuring the
growth and destruction rate in cosmological simulations.

The shapes of voids offer a particularly interesting cosmological
probe, whether by their distribution~\citep[e.g.][]{Bos2012, Li2012}
or via an application of the Alcock-Paczynski
test~\citep{Alcock1979, Ryden1995, LavauxGuilhem2011, Sutter2012b, Sutter2014b}.
However, these tests rely on the assumption that the void identified
in a galaxy survey corresponds to a physical underdensity in the
dark matter. While this is largely an issue of sparsity and
galaxy bias~\citep{Sutter2013b}, the watershed technique may
spuriously merge voids even in the dark matter.
These voids will erroneously
appear as larger voids that are not completely
empty and thus have suspect shapes. We can use a detailed merger history
to identify such suspect voids.

Recently~\citet{Hamaus2013}
pointed out that for a given
tracer population there exists a \emph{compensation scale}, 
where the void-matter bias is identically zero.
Below this scale, voids generally collapse due to 
their surrounding overdense walls, while above this scale
voids tend to continue expanding~\citep{Ceccarelli2013}. 
However, these results are based
on studies of the velocity profiles and clustering statistics
at fixed time. 
Only by tracing the evolution --- and thereby studying the 
dynamics --- of voids could one accurately 
examine the properties of voids in relation to such a compensation scale.

Finally, the growth and merger rates of voids are potential
cosmological probes, analogous to the growth rate of cosmic structure.
The nature of modified gravity and fifth forces can leave fingerprints on the evolution of the void population at high redshift, potentially constraining the properties of dark energy.

Unfortunately, to date this remains a largely
 unexplored topic. Most early studies of voids in simulations focused
on visual identification and characterization~\citep[e.g.,][]{White1987}.
For example, the pioneering works
of~\citet{Dubinski1993}, which discussed the process of
void merging, and~\citet{vandeWey1993}, which first noted
the hierarchical nature of void buildup, were entirely based on
visually examining thin slices of $N$-body simulations.
More recent and more sophisticated
analyses have focused on void interiors~\citep[e.g.,][]{Gottlober2003,
Goldberg2004,Aragon2012, Neyrinck2013} or on statistics at a fixed
time such as those discussed above.

In this work we present a comprehensive study of the formation, subsequent evolution, and destruction of voids. We use techniques adapted from building halo
merger trees~\citep{Srisawat2013}
to follow individual voids across
cosmic time. This approach allows us to measure their formation
time, identify when mergers occur, track their movement and growth,
and, when it does happen, record their time of collapse. We translate this
information into rates and correlate these rates with void size,
which can then inform theoretical and observational
results.

In the following section we review our simulation setup, void finding
approach, merger identification technique, and some definitions to
be used throughout the work. In Section~\ref{sec:formation} we focus on the
formation time of voids, followed by a discussion
in Section~\ref{sec:growth} of their growth and merger histories.
In Sections~\ref{sec:movement} and~\ref{sec:destruction}
we present an analysis of void movement and destruction rate
over cosmological time, respectively.
Finally, we conclude in Section~\ref{sec:conclusions}
with a brief discussion of implications for theoretical modeling of voids
and directions for future work.

\section{Numerical Approach}
\label{sec:numerical}

\subsection{Simulation}

We study voids forming in a single cosmological dark matter simulation run using the \textsc{Gadget-3} $N$-body 
code~\citep{Springel2005} with initial conditions drawn from the WMAP-7 cosmology~\citep{Komatsu2011}.
Voids are identified in 62 snapshots from redshift 0 to $\sim$30.
The snapshots are evenly spaced in $\ln a$, where $a$ is the scale
factor. The simulation contains $270^3$ particles in a box of comoving length $62.5$ $h^{-1}$Mpc, giving a dark matter
particle mass of $9.31\times10^8 \hmsol$. 
This combination of box size and number of particles gives a mean 
interparticle spacing of $\sim 0.25$ \hmpc. Since we will only study of 
voids of effective radius $1$ \hmpc~(see below), this provides sufficient 
resolution for even the smallest voids and allows us to examine 
several thousand voids. Increasing resolution 
or box size would give 
us access to even more voids, but the analysis of~\citet{Sutter2013b} 
shows that voids are self-similar up to a scale 
of $\sim100$\hmpc~in a \lcdm~universe: 
studying this distribution of voids 
gives us a fairly representative picture.

For more simulation details see \cite{Srisawat2013}.

\subsection{Void Finding}
We identify voids with a heavily modified and extended version of
\textsc{zobov}~\citep{Neyrinck2008} 
called \textsc{VIDE}~\citep{Sutter2014c}.
\textsc{VIDE} creates a Voronoi tessellation of the tracer particle
population and uses the watershed transform to group Voronoi
cells into zones and subsequently voids~\citep{Platen2007}.
By implicitly 
performing a Delauney triangulation (the dual of the Voronoi 
tessellation), \textsc{VIDE} assumes constant density across the volume 
of each Voronoi cell, which sets the smoothing scale for the continuous
field necessary to perform the 
watershed transform. There is no additional smoothing.

The algorithm proceeds by first grouping adjacent Voronoi 
cells into {\emph zones}, which are local basins.
Next, the watershed transform merges zones into voids 
by examining the density barriers between them and joining them 
together to form ever-larger agglomerations. 
We impose a density-based threshold within \textsc{VIDE} where
adjacent zones are only added to a void 
if the density of the wall between them is
less than $0.2$ times the mean particle density.
Derived from the characteristic void nonlinearity density 
level~\citep{Platen2007}, this prevents voids from expanding deeply into overdense structures and 
limits the depth of the void hierarchy~\citep{Neyrinck2008}.
However, this does not place a restriction on the density of the 
initial zone, and in principle a void can have any mean density.

The watershed transform identifies catchment basins
as the cores of voids and ridgelines, which separate the flow
of water, as the boundaries of voids.
In sum, we identify voids as depressions in the tracer density; voids are
non-spherical aggregations of Voronoi cells that share a common
basin and are bounded by a common set of higher-density walls.

\begin{figure} 
  \centering 
  {\includegraphics[type=pdf,ext=.png,read=.png,width=\columnwidth]{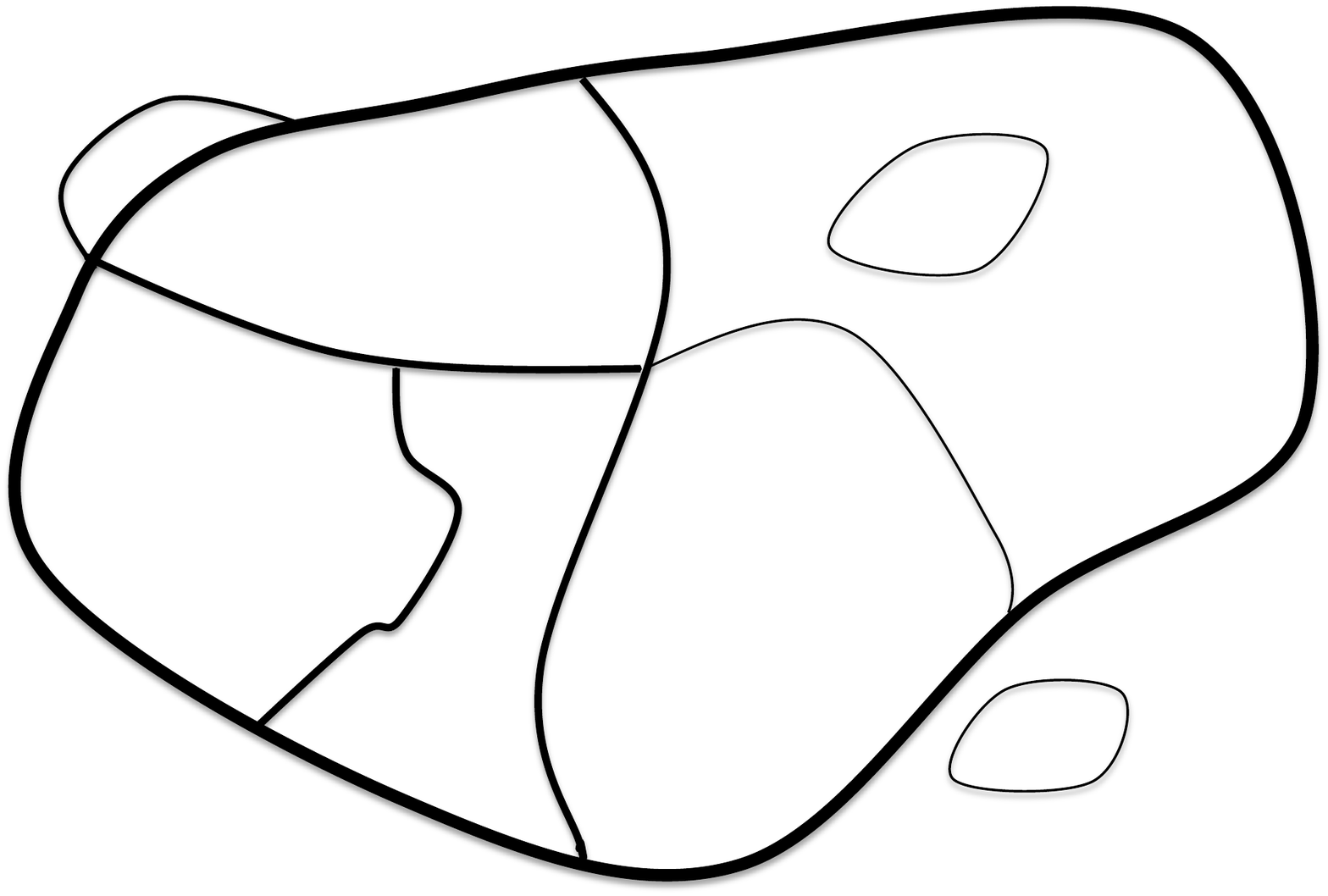}}
  {\includegraphics[type=pdf,ext=.pdf,read=.pdf,width=\columnwidth]{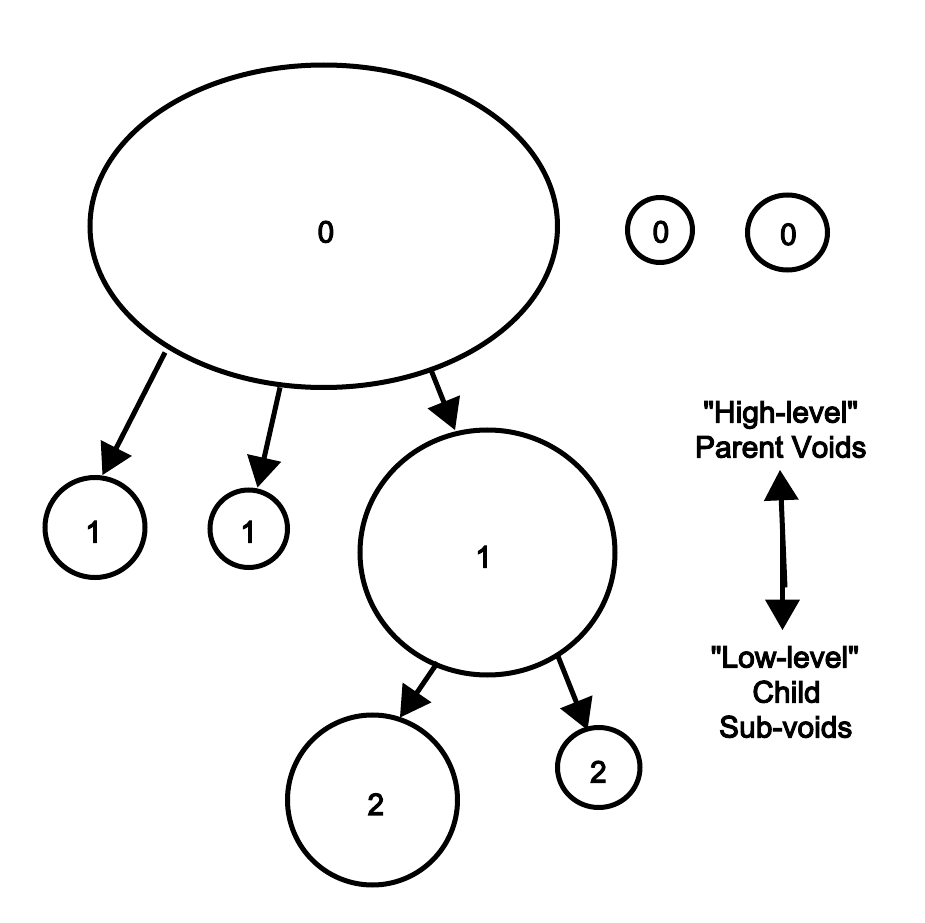}}
  \caption{
           A cartoon of the assembly of the void hierarchy. The top panel 
           shows ridgelines with line thickness proportional to density.
           The bottom panel shows the tree derived from such a collection
           of voids, with the tree level of each void indicated.
          }
\label{fig:hierarchy}
\end{figure}

These operations allow the construction of a nested hierarchy of
voids~\citep{LavauxGuilhem2011, Bos2012}: we identify the initial 
zones as the deepest voids, and as we progressively merge voids 
across ridgelines we can identify super-voids. There is no unique 
definition of a void hierarchy, and we take the semantics 
of~\citet{LavauxGuilhem2011}: a parent void contains all the zones 
of a sub-void plus at least one more. All voids have only one 
parent but potentially many (or no) children, and the children 
of a parent occupy distinct subvolumes separated by low-lying 
ridgelines. Figure~\ref{fig:hierarchy} shows a cartoon of this 
void hierarchy construction.
For visualizations of watershed voids, we refer the reader to 
\citet{Platen2007},~\citet{Neyrinck2008},~\citet{Colberg2008},
~\citet{Bos2012}, and~\citet{Sutter2012a}, 
Also,~\citet{Aragon2010} shows the nested hierarchy of voids 
using different, but related, assembly criteria.

Our simulation gives a mean particle spacing 
$\bar{n}^{-1/3} \approx 0.25$~\hmpc, which sets a lower size 
limit of the detectability of voids due to shot noise. 
For this work we will 
study all voids with effective radius $R_{\rm eff} > 1$~\hmpc.
We define the effective radius as
\begin{equation}
  R_{\rm eff} \equiv \left( \frac{3}{4 \pi} V \right)^{1/3},
\end {equation}
where $V$ is the total volume of the Voronoi cells
that contribute to the void.
Note that this cut is at a much larger radius than used in 
previous analyses with \textsc{VIDE}~\citep[e.g.,][]{Sutter2013c}. 
While voids near the mean particle separation do not appear to be
simple Poisson noise~\citep{Hamaus2014}, small-scale 
fluctuations in the density field can still give rise to occasional 
spurious features which are not filtered out, leading to an incomplete
and inaccurate
picture of the void population at these scales. 
However, at four times the mean particle separation, the abundance 
of voids appears complete~\citep[e.g.,][]{Sutter2013a}, 
the void properties are convergent to higher resolutions~\citep[e.g.,][]{Sutter2013b},
and the contamination by Poisson fluctuations is 
exponentially diminished~\citep{Neyrinck2008,Sutter2013d}.
Thus, for our study this criterion gives a rather robust 
picture of the void population.
We do not impose any other cuts based on density contrast 
or minimum density --- we wish to see if marginal voids have 
similar histories as deeper underdensities.

Additionally, for the analysis below we need to define a center for each
void. For this work we take the \emph{macrocenter}, or volume-weighted
center of all the Voronoi cells in the void:
\begin{equation}
  {\bf X}_v = \frac{1}{\sum_i V_i} \sum_i {\bf x}_i V_i,
\label{eq:macrocenter}
\end{equation}
where ${\bf x}_i$ and $V_i$ are the positions and Voronoi volumes of
each tracer $i$, respectively.

\begin{figure} 
  \centering 
  {\includegraphics[type=png,ext=.png,read=.png,width=\columnwidth]{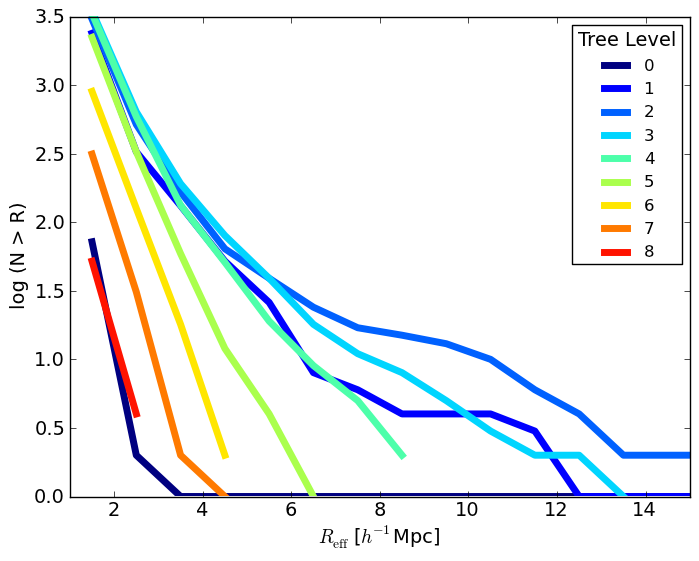}}
  \caption{
           Cumulative number functions for voids in different levels of the 
           hierarchy. Tree level 0 (dark blue) are the topmost parent voids,
           and tree level 8 (red) are subvoids deepest in the hierarchy. 
           Though higher-level
           voids tend to be larger, they span a broad range of sizes.
           At the topmost level, there are only a few small ``field'' 
           (i.e., childless) voids and some voids that span nearly the entire
           simulation volume.
           We do not show these largest voids so that we may 
           highlight the relative differences of the remaining hierarchy 
           levels.
          }
\label{fig:treedist}
\end{figure}

Figure~\ref{fig:treedist} shows the cumulative void number function 
for all voids in the final $a=1.0$ snapshot, organized by level 
in the hierarchy. Note that these numbers functions do not turn over at 
small radii, as predicted by~\citet{Sheth2004}, 
since we are plotting the cumulative, rather than differential, function, 
and we are only counting voids well above the resolution limit. 
At the top-most parent root level (Tree Level 0) there are only a few 
small field voids and the largest voids in the simulation. 
As we go deeper into the hierarchy, we see increasingly 
smaller voids. 
We construct the  
void tree such that each void has only a single parent (or no parents at all)
and can potentially have many children. One void is a parent of
another if it shares all zones of the child plus at least one more.
Parents can then become children of even larger super-voids. Without any
density thresholds, there will be a single void that encompasses the entire
simulation volume. 
However, since we do apply a density threshold, we have multiple root
voids.
The relatively small simulation box prevents us from examining the very 
largest voids; however, the large voids that are discovered in this 
box are representative of the voids found in larger simulations 
and galaxy surveys.

\subsection{Merger Identification}

To match voids from one snapshot to another, we use the tree building routine that is part of the publicly available \velociraptor\ (aka \textsc{STF})
package\footnote{{https://www.dropbox.com/sh/177zo6q3qk5pdkz/T0V0eseLZu}}.
The \velociraptor\ tree builder code is a particle correlator: it takes two particle ID lists (named $A$ and $B$) and for each object in list $B$ 
identifies those objects in list $A$
(i.e., in the previous snapshot) that have particles in common. This first step 
produces a graph mapping the connections between objects rather than an actual 
progenitor tree. To produce a tree the algorithm calculates the merit of each connection:
\begin{equation}  
  \mathcal{M}_{A_i B_j} = N_{A_i\cap B_j}^2/(N_{A_i} N_{B_j}),
\end{equation}
where $N_{A_i\cap B_j}$ is the number of shared particles between the 
object $i$ in catalog $A$ and object $j$ in catalog $B$, 
and $N_{A_i}$ and $N_{B_j}$ are the number of particles in 
object $i$ in catalog $A$ and $j$ in $B$, respectively. 
Between two snapshots, the unique main progenitor is the object 
that maximises the merit. This strategy has proven successful in building {\em halo} merger trees \cite[see for instance][]{Srisawat2013}.

Note that while technically we are constructing bi-directional 
\emph{graphs}, we will see that we are justified in calling these \emph{trees} 
since the void evolution is surprisingly simple.

However, unlike halos, which are defined by the particles they are composed of, voids are defined by the empty spaces between particles. Ideally we would correlate \emph{volumes} rather than \emph{particles}. 
However, this is computationally expensive and fraught with difficulties: it would require modeling the Voronoi volume around each particle and 
making arbitrary decisions on when one particle's volume is correlated 
with another. Instead of trying to determine the overlap in 
volume, we simply use the volume associated with each particle as determined 
by \textsc{VIDE}, $v_l$, to weight the merit function. 
Thus the modified merit function becomes
\begin{equation}
  \mathcal{M}_{A_i B_j} = \tilde{V}_{A_i\cap B_j}^2/(V_{A_i} V_{B_j}),
\end{equation}
where the total volume of a void is $V=\sum_l^N v_l$, and the shared volume is $\tilde{V}_{A_i\cap B_j}^2=\sum_{l}^{N_{A_i\cap B_j}} v_{l,A_{i}} v_{l,B_{j}}$. This sum is over all shared particles, 
with each particle weighted by the volume associated with it 
in catalogs $A$ and $B$. In order to qualify as a progenitor 
the compared voids must share at least 10 particles, though 
changes to this value do not produce significantly different results.

We note that even a simple particle based 
approach is justified as even our smallest voids contain hundreds of particles. While the cores of voids are empty, there are sufficient numbers of particle distributed throughout the remaining volume that particles can be used as a proxy for the void volume. The volume weighting scheme used here simply reduces instances where a void in catalog $A$ shares particles with several voids 
in catalog $B$. The volume weighted merit function chooses the progenitor which minimises the volume fluctuations on a particle by particle basis and for the void as a whole.

In short, to construct a progenitor history for each void we find the void in
the previous snapshot that shares the most number of particles, weighted by 
volume. This is called the \emph{main progenitor}.
We discuss the evolutionary chain of these main progenitors
for the remainder of this work when we examine the movement and 
growth of a void.

\section{Void Formation}
\label{sec:formation}

\begin{figure*} 
  \centering 
  {\includegraphics[type=png,ext=.png,read=.png,width=0.32\textwidth]{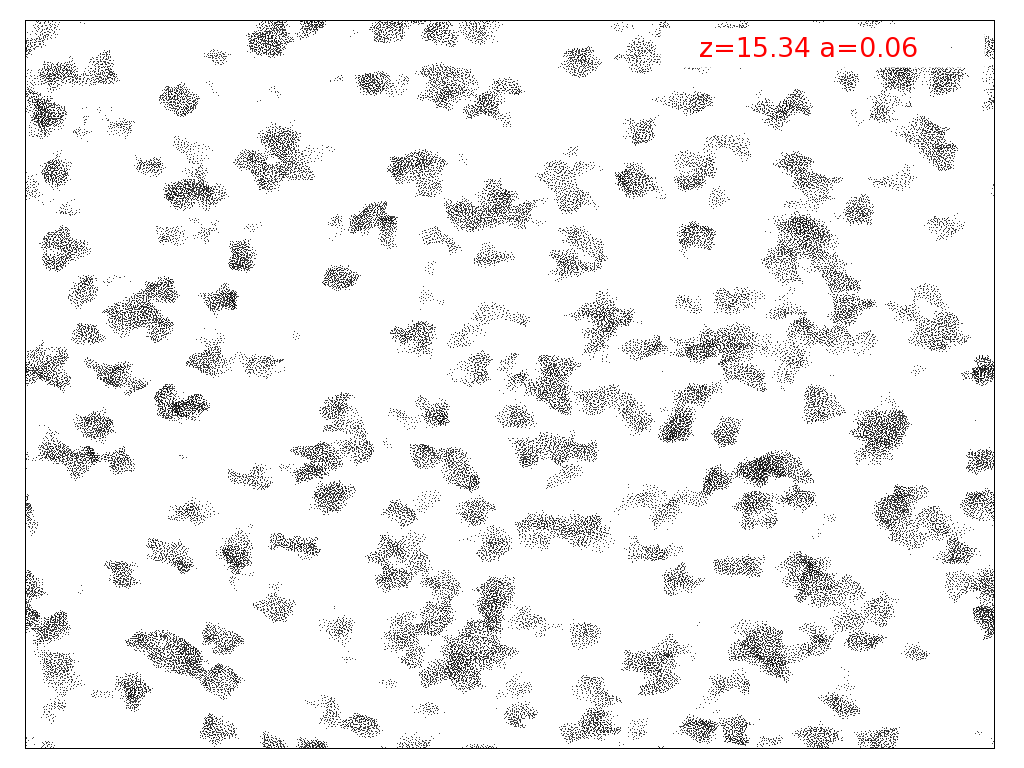}}
  {\includegraphics[type=png,ext=.png,read=.png,width=0.32\textwidth]{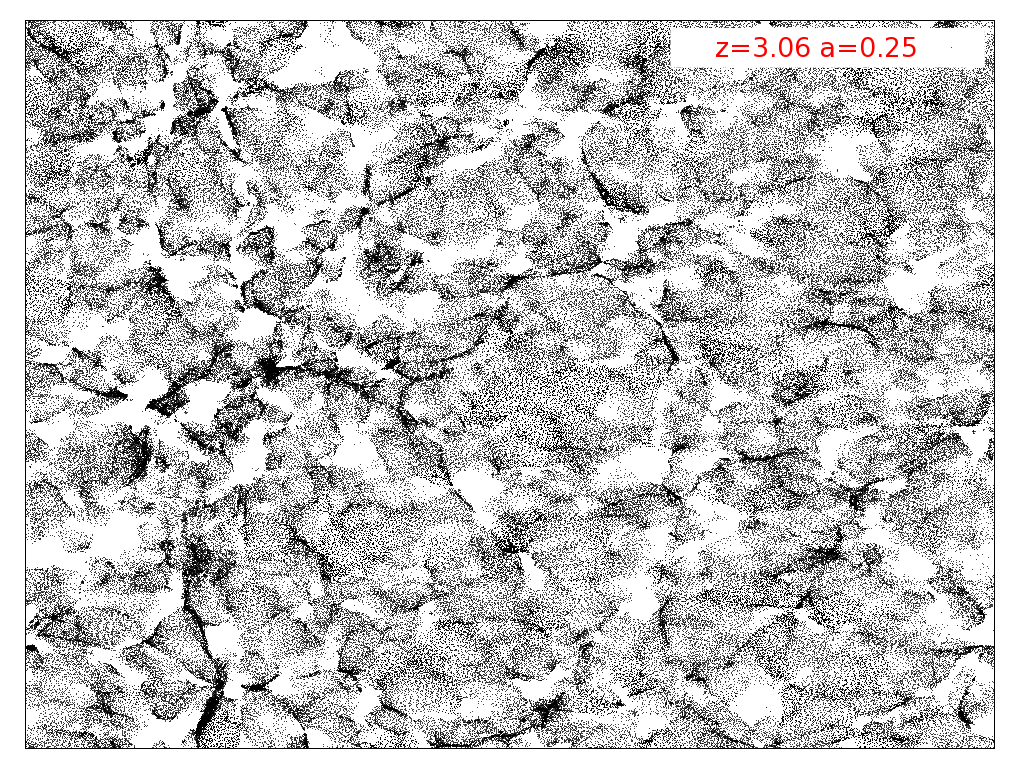}}
  {\includegraphics[type=png,ext=.png,read=.png,width=0.32\textwidth]{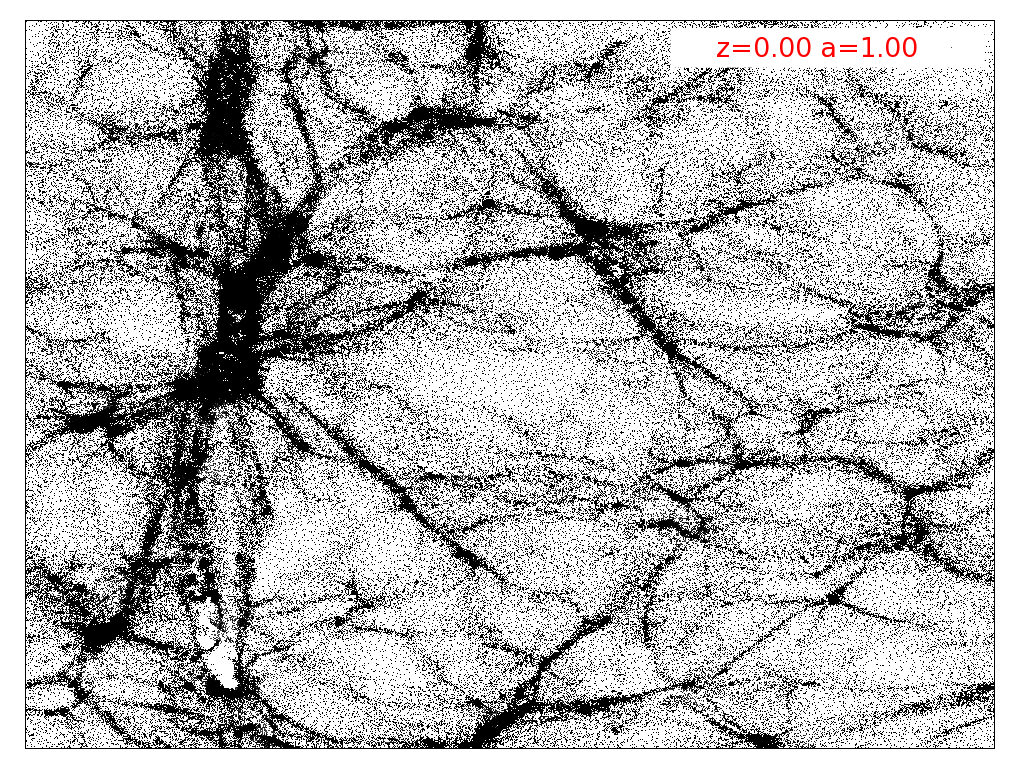}}
  \caption{
           A visual impression of the buildup of voids. Shown are thin slices
           of the particle distribution at various simulation snapshots. 
           Only void member particles are shown. Slices are at scale 
           factors $0.1$ (left), $0.25$ (middle), and $1.0$ (right). 
           Around $a=0.23$, filaments and walls become dense enough and the 
           underdensities clear enough to support
           the formation of larger voids. While the parent voids  
           gently expand, the formation and merging of subvoids continues.
           Watershed void finders include as void members all particles 
           within the highest-density ridgeline; hence at late times almost 
           all but the highest-density particles are included in voids.
          }
\label{fig:mergertrees}
\end{figure*}

We begin with visual examination of voids as a 
function of scale factor, shown in Figure~\ref{fig:mergertrees}.
This figure shows slices of the dark matter particles that are 
identified as belonging to voids. Initially there are only small,
isolated voids just above the minimum size threshold.  
Since we do not apply a density criterion, these are shallow 
basins that will eventually empty out (as seen in 
~\citealt{vdW2004}). 
These depressions in the initial density field then begin 
to expand, with small basins quickly merging with larger basins. 
As the walls and filaments begin to coalesce around $a=0.2-0.4$, 
a complex void hierarchy begins to form as subvoids group into larger 
parent points. 
At this epoch the density contrast in the 
large scale structure becomes high enough for our void finder to 
identify the multi-level basins as distinct voids. 
At late times, the larger voids simply 
grow and expand into their local environment, and since the watershed 
method includes all particles within the ridgeline as void members, 
we see very few gaps (e.g., the dense halos) in the void particle 
distribution.

The shallow basins at early times could be considered as ``proto-voids''.
To separate these from mature voids one could define a density 
threshold, as is done for halos~\citep{Sheth2004}. 
While linearly-extrapolated initial conditions do not always 
clearly map to final void states~\citep{Sahni1996}, there is 
evidence that void \emph{properties} 
exhibit only linear and quasi-linear behavior. 
For example, the recent work of~\citet{Hamaus2014} showed that density and velocity profiles of voids, while each individually described by non-linear functions, can be related by linear perturbation theory very accurately.
Also,~\citet{Lavaux2010} showed 
that it is possible to map the shapes of voids in Lagrangian initial 
conditions to the late-time Eurlerian object.
Thus, there is no clear 
distinction between early- and late-time voids: there is a one-to-one 
mapping from the initial to the final void state.

\begin{figure}
  \centering
  {\includegraphics[type=png,ext=.png,read=.png,width=\columnwidth]{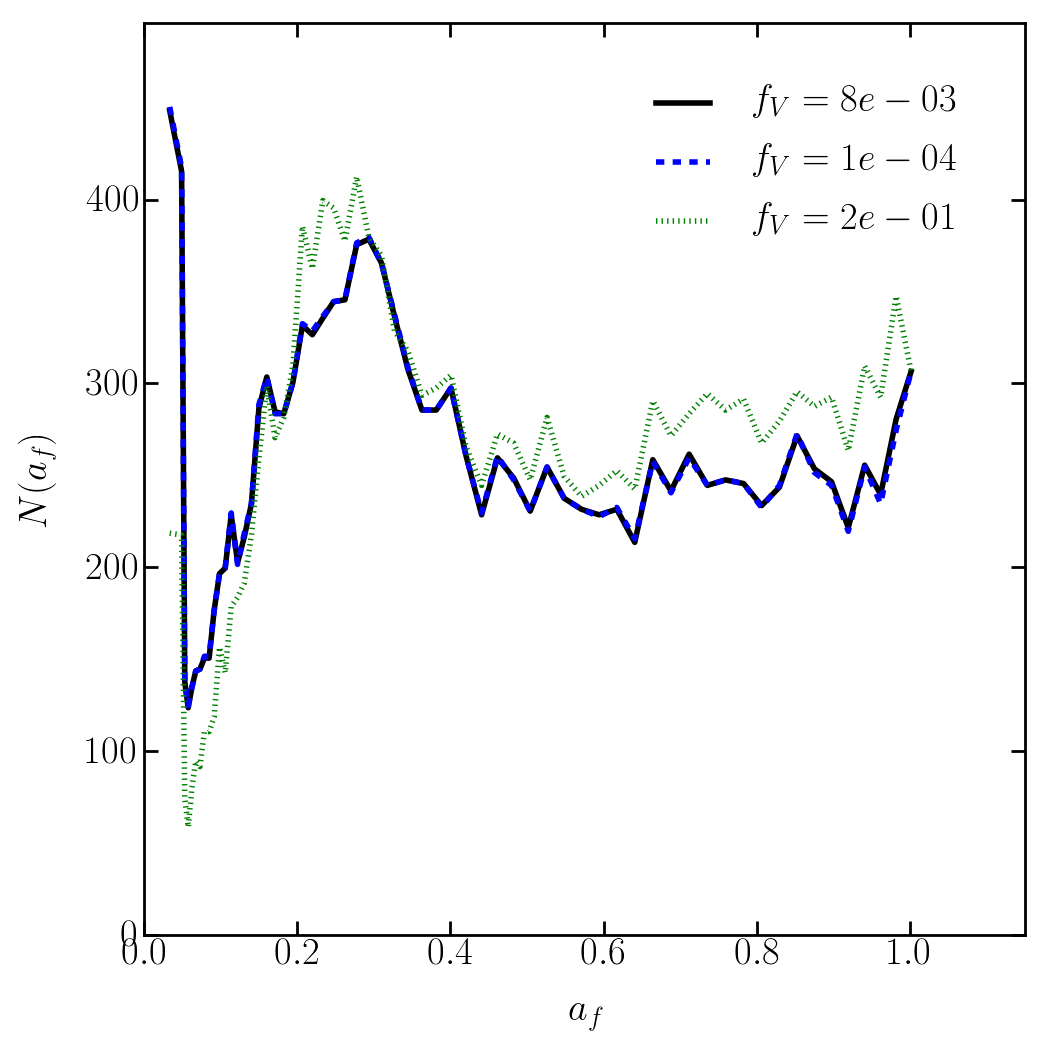}}
  \caption{
           Distribution of scale factors of
           formation for voids that persist to $a=1.0$.
           The peak around $a=0.2-0.4$ corresponds to a 
           significant increase in the depth
           of the void hierarchy.
          }
\label{fig:tform}
\end{figure}

To clarify this mapping, 
Figure~\ref{fig:tform} shows the formation scale factor $a_f$ for each void
identified at the present day. We define $a_f$ as the scale factor
at which the void reaches a fraction $f_V=V(a_f)/V_o$ of its current volume, $V_o$. Since voids grow {\em and shrink}, defining a formation time is not trivial. We set $a_f$ to the time at which the void's volume drops below $f_V V_o$ for the {\em first} time as we go back along a void's history. 
We indirectly investigate our definition of a formation time by examine several values of $f_V$. We use $10^{-4}$, $8 \times 10^{-3}$,
and $0.2$, which appear chosen arbitrarily but 
pertain to specific increases in void size by corresponding to fractional 
radii of $0.01$, $0.05$, and $0.1$, respectively. All 
three values give nearly identical results: some voids
have persisted since the beginning of the simulation, and others have
only appeared recently. 
As expected, with larger values of $f_V$ the distribution skews to 
later times (note the much smaller value of the green line at 
$a=0.0$), but this is surprisingly insignificant: once a void 
is detected, it has essentially its present-day volume.
There is also an increase in the formation time just before $a=1.0$. 
While this may be due to numerical effects, it is not significantly 
different than the $a_f>0.4$ fluctuations.

There is a noticeable spike at
$a\approx0.3$. This coincides with the initial growth of the void hierarchy;
prior to this time there are only isolated voids. Thus this time
indicates the appearance of significant structural hierarchy in the
cosmic web. 
It also coincides with the appearance --- though not domination --- 
of dark energy: in these epochs, $\Omega_\Lambda \sim 0.1 \Omega_M$. 
Thus a large void population forms at these scale factors 
through a
combination of the crystallization of the cosmic web, allowing 
basins to be identified, and 
the emergence of dark energy, which shuts off significant 
continued void production.
Since all chosen values of $f_V$ give nearly the same results,
voids reach their present-day volume quickly --- essentially when
they are first able to be identified as voids --- and do not
grow much. We will return to this subject later.

\begin{figure}
  \centering
  {\includegraphics[type=png,ext=.png,read=.png,width=\columnwidth]{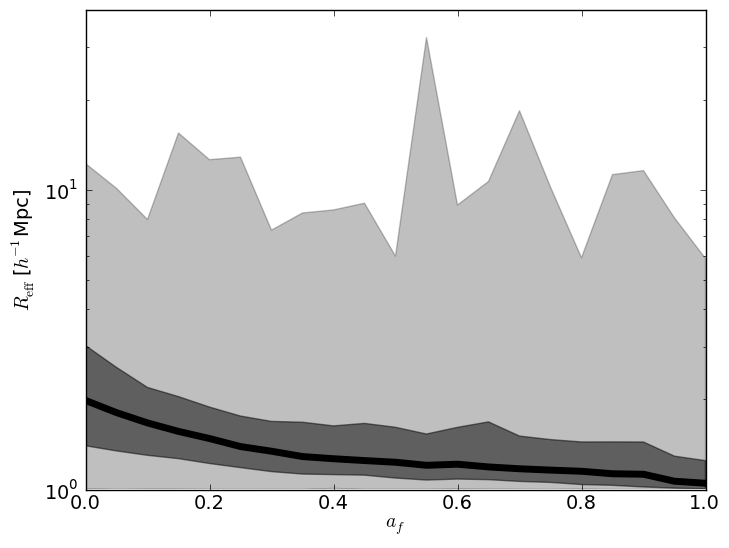}}
  \caption{
           Distribution of $a=1$ void sizes as a function of their formation
           scale factor $a_f$. The black line is the median in bins of
           width $\Delta a_f=0.05$, the dark bands are the inner 68\% of 
           the binned distribution, and light grey bands
           are the bin extrema. 
          }
\label{fig:rsize}
\end{figure}

In Figure~\ref{fig:rsize} we examine the distribution of void sizes
as they appear in the simulation. Throughout most of cosmic history,
the median void formation size is centered on $2$~\hmpc, and a 
small population of medium-scale $10-15$~\hmpc~voids continually appears.
The median void formation drops to nearly $1$~\hmpc~by the 
present day, since newer voids can only occupy smaller niches 
in the cosmic web adjacent to larger, expanding neighbor voids.
The very largest voids appear at any times, as the walls between
mid-scale voids empty out and the voids are joined into larger
super-voids. The fact that there is no noticeable peak 
formation time for voids of a specific size but there is a peak in 
the formation \emph{time} of voids indicates that the
level of these voids in the hierarchy changes with time.
Although we are showing the present-day $a=1$ void sizes, the fact 
that voids do not grow much over time indicates that this is also 
essentially their size at formation.

\section{Void Mergers \& Growth}
\label{sec:growth}

\begin{figure*} 
  \centering 
{\includegraphics[type=pdf,ext=.pdf,read=.pdf,width=0.33\textwidth]{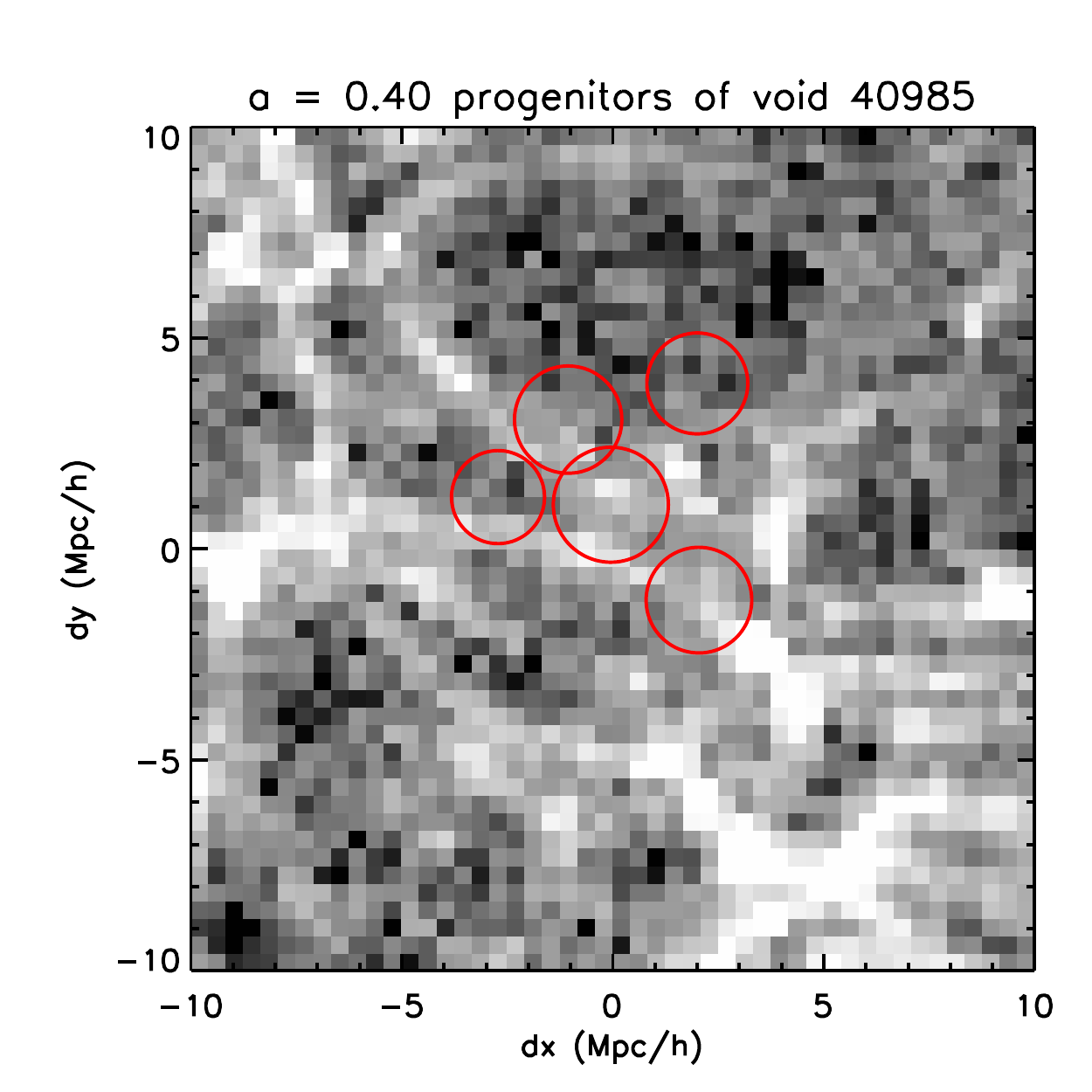}}
  {\includegraphics[type=pdf,ext=.pdf,read=.pdf,width=0.33\textwidth]{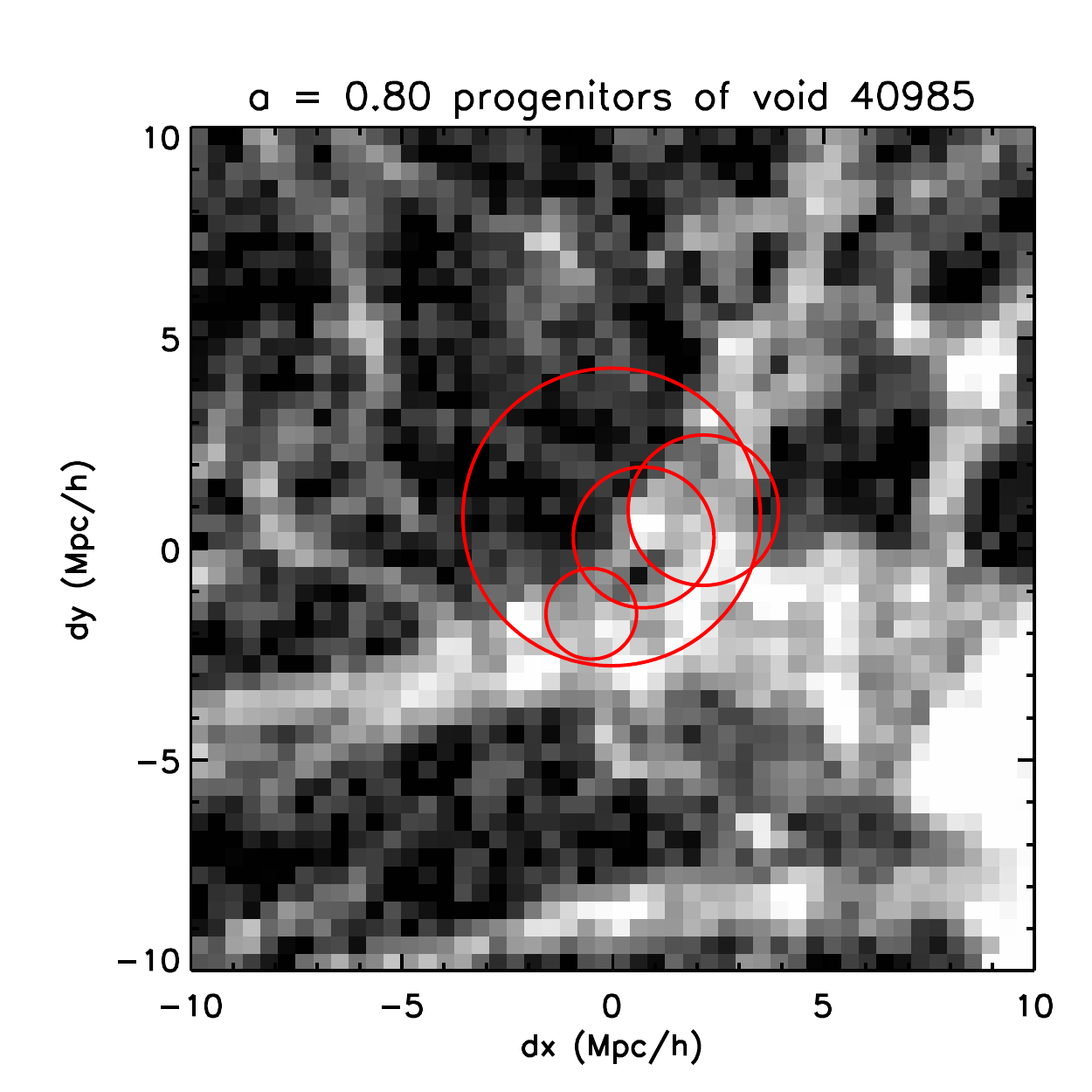}}
  {\includegraphics[type=pdf,ext=.pdf,read=.pdf,width=0.33\textwidth]{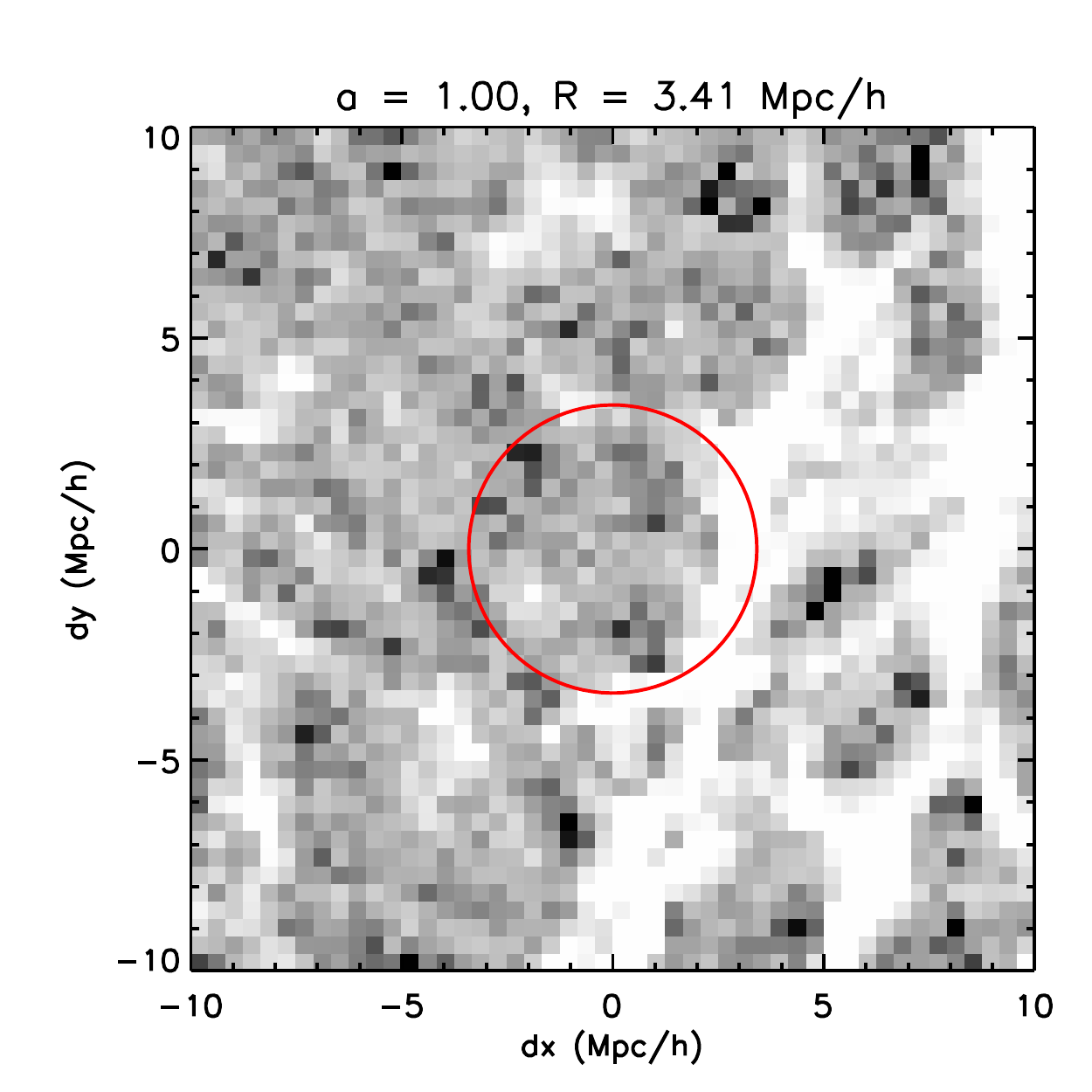}}
  \caption{
           Evolution of progenitor voids.
           We show thin slices through the dark matter density with voids superimposed
           on top. Voids are represented as circles with radii equal to 
           $R_{\rm eff}$. Slices are arranged from early (left) to late (right) 
           times and trace the evolution of a singe void to 
           highlight different void merger histories. 
           We only show progenitors of the final $a=1.0$ void. 
            To increase the density contrast, each panel is scaled such that black and white are the minimum and maximum density of the cells shown, respectively. The density is constructed using cloud-in-cell weighting 
            and the shading is scaled according to $\log{(1+\delta)}$.
           Projection effects lead to void centers occasionally appearing 
           to lay on top of filaments.
           Several processes are highlighted by this evolution, including 
           void merging as barriers dissolve, and the formation of 
           a void hierarchy as the larger parent void annexes smaller,
           but distinct, subvolumes.
          }
\label{fig:slices}
\end{figure*}

We use Figure~\ref{fig:slices} to show
an example of a void merger history. In this figure we show 
slices of the dark matter density centered on a  representative
void at $a=1.0$. We then show the same slice at the same position 
at two other scale factors, $0.8$ and $0.4$. In these additional slices 
we show the progenitors of the present-day void.
While the voids in general have complex shapes due to the nature of the 
watershed (see, for example,~\citealt{Sutter2012a}) we represent them 
here as simple circles with radii equal to the effective radius
$R_{\rm eff}$; our purpose with this illustration is to simply provide 
a rough guide to the eye of void location and size relative to the 
cosmic web, not to examine the detailed impact on void shapes.

We choose this void to highlight two distinct processes of void evolution.
The first, pure merging, occurs when high-density barriers between two 
voids completely dissolve due to outflows. When the barrier becomes 
too low ($<0.2 \bar{\rho})$, the separate voids become indistinguishable 
from each other, forming a single larger void. However, as we will 
discuss below, this is a very rare process. Instead, what more frequently 
occurs is \emph{annexation} of subvoids as a larger parent void
assembles, generating a hierarchy. 
In this picture, a subvoid undergoing annexation retains its detectability 
and individual identity, and comprises only a portion of the volume of a 
larger parent (Figure~\ref{fig:hierarchy}). In contrast, a \emph{merging} 
subvoid loses its identity and is no longer separately detectable.

\begin{figure} 
  \centering 
  {\includegraphics[type=png,ext=.png,read=.png,width=\columnwidth]{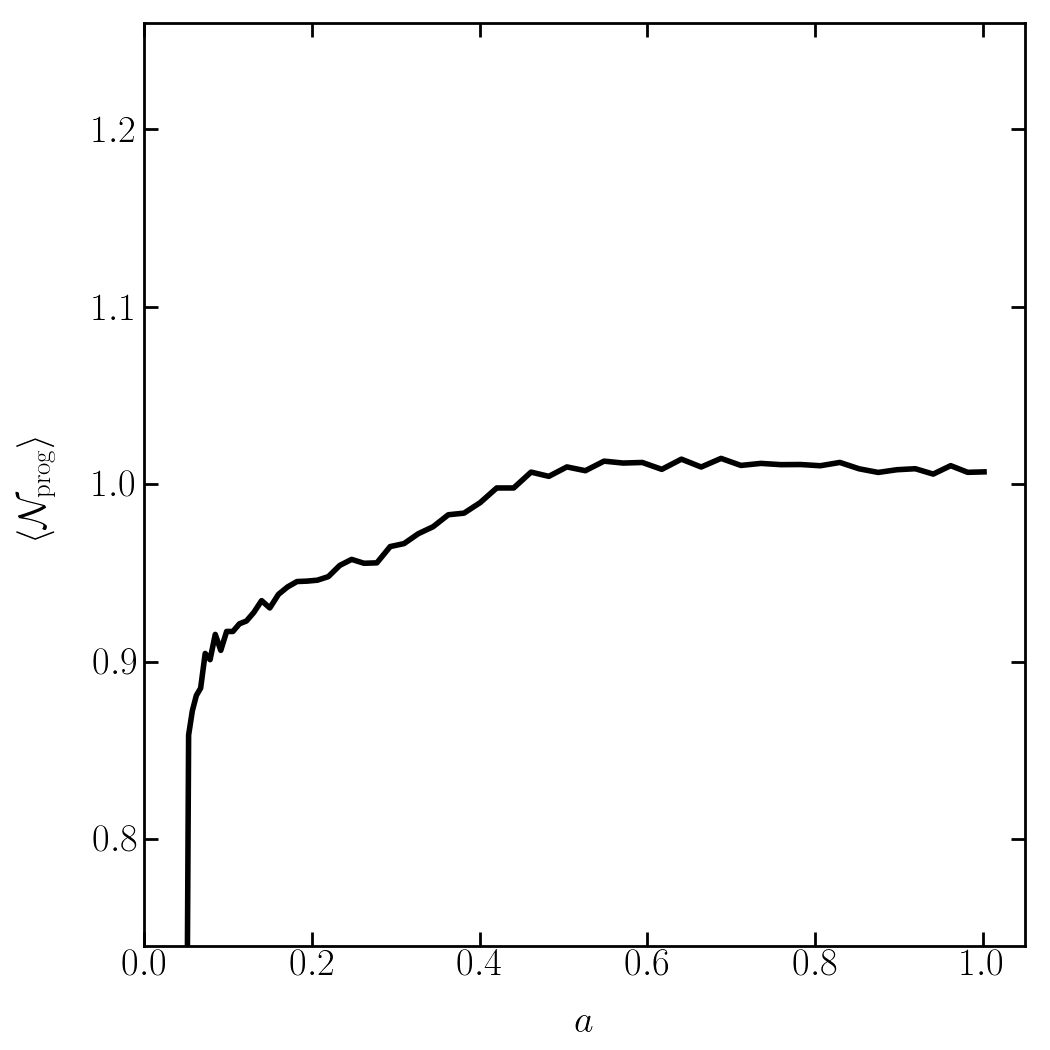}}
  \caption{
           Average number of progenitors as a function of scale factor 
           for voids at $a=1.0$. We calculate this quantity by 
           counting the total number of progenitors for all present-day
           voids and dividing by the number of present-day voids. Hence, 
           this quantity becomes less than one as the lines of descent 
           for individual voids end.
          }
\label{fig:avenprog}
\end{figure}

For each progenitor tree leading up to each present-day void, 
we can track the total number of progenitors; in other words, the width 
of the merger tree. In Figure~\ref{fig:avenprog} we plot the 
average number of progenitors as a function of the scale factor for all present-day voids,
\begin{align}
  \left< N_{\rm prog} \right>(a)=\frac{1}{N_{{\rm voids},o}}\sum_i N_{{\rm prog},i}(a).
\end{align}
Here the sum is over all progenitors at a time $a$. 
The fact that the mean number of progenitors is barely greater than 
one indicates that almost all voids follow only a single line 
of descent and experience very few mergers. 
The number of progenitors begins to decrease at 
$a=0.4$, which coincides with the end of the peak formation 
time (Figure~\ref{fig:tform}) and the relative lack of 
new voids after that time. 

\begin{figure} 
  \centering 
  {\includegraphics[type=png,ext=.png,read=.png,width=0.48\textwidth]{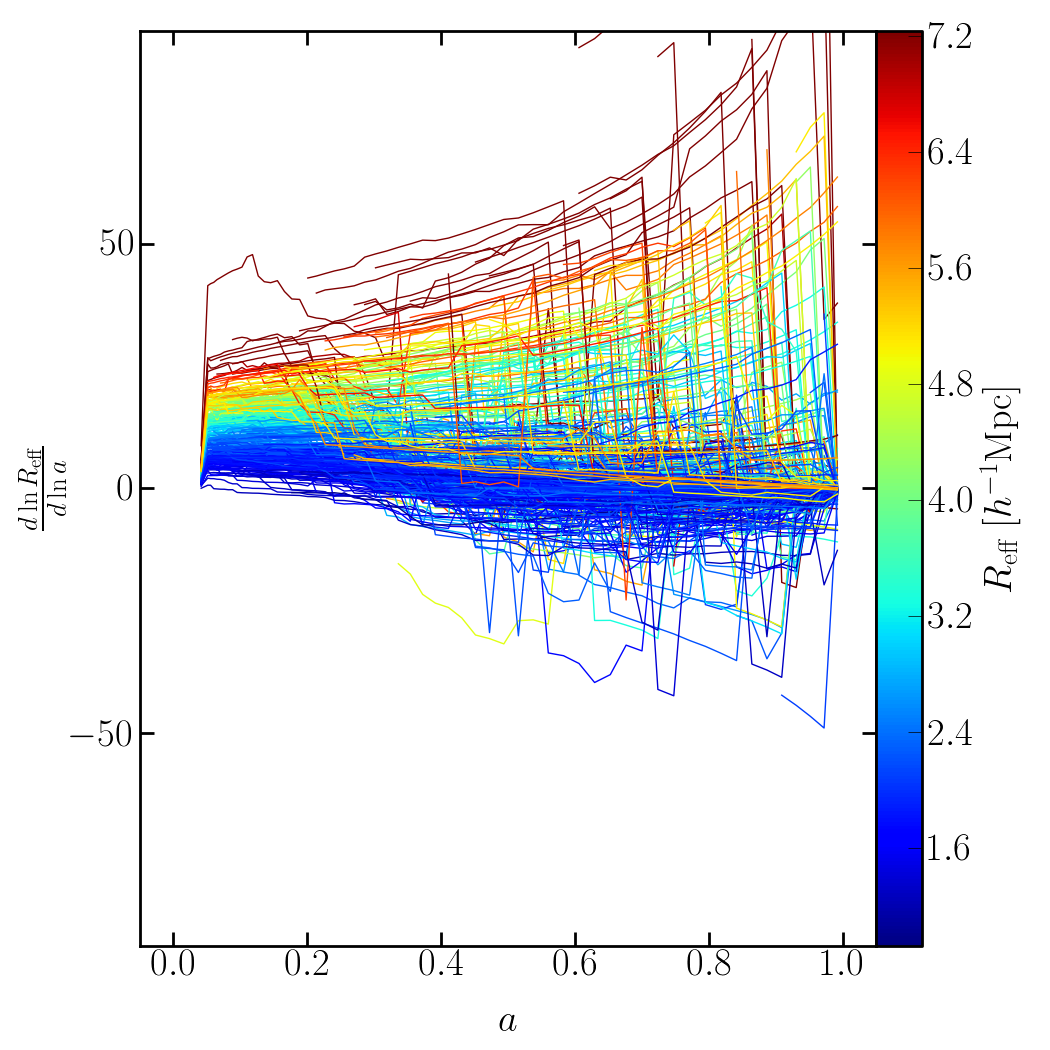}}
  \caption{
           Growth rate as a function of scale factor for each void 
           surviving at $a=1.0$. Individual lines are colored by void size, 
           from the smallest (blue) to largest (red). 
           The radii listed are the final $a=1.0$ size.
          }
\label{fig:growth}
\end{figure}

In Figure~\ref{fig:growth} we show the growth rate history for every void 
in the simulation as a function of scale factor $a$,
\begin{align}
\frac{d \ln R_{\rm eff}}{d \ln a}=\ln(R_{i+1}/R_{i})/\ln(a_{i+1}/a_{i}),
\end{align}
where the change is calculate between the $i^{\rm th}+1$ and $i^{\rm th}$ snapshots.
We plot a line for each individual void. As expected, the voids with 
the highest growth rates are the largest; these are the supervoids 
that form from the rapid merger of subvoids as the basin 
empties out of substructure.
Since they form in deeply-underdense environments 
with little substantial structure surrounding them, they 
act as miniature universes with $\Omega_{\rm tot} < 1$
~\citep{Goldberg2004}. For these 
larger voids there are a few steep changes as they merge with a 
smaller void or fragment into smaller progenitors. 
Even though the merger or fragmentation ratio is small in terms of volume,
it can impact the instantaneous growth rate from one snapshot to another.

The small- and medium-scale voids show remarkably steady growth histories, 
with very few strong deviations. Even though some of these voids 
do merge, they tend to just absorb their subvoids, so the overall 
volume gained is small.
Interestingly, there is
 a population of collapsing voids: these are the 
voids located in overall overdense regions. This is the ``void-in-cloud'' 
phenomenon of~\citet{Sheth2004}. Even though there are a few small 
voids with discontinuous merger histories, almost all the small voids 
are either gently expanding or contracting. 

\begin{figure*} 
  \centering 
  {\includegraphics[type=png,ext=.png,read=.png,width=0.48\textwidth]{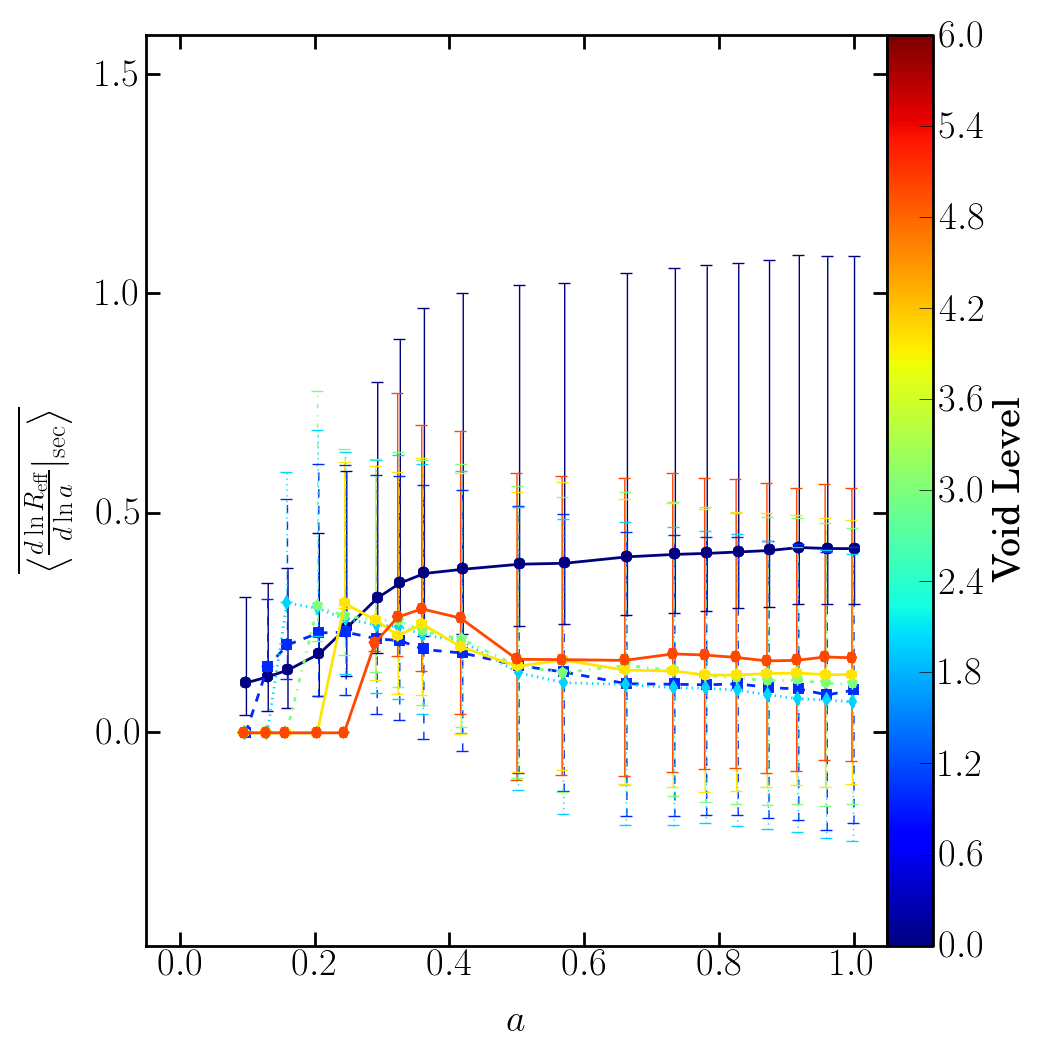}}
  {\includegraphics[type=png,ext=.png,read=.png,width=0.48\textwidth]{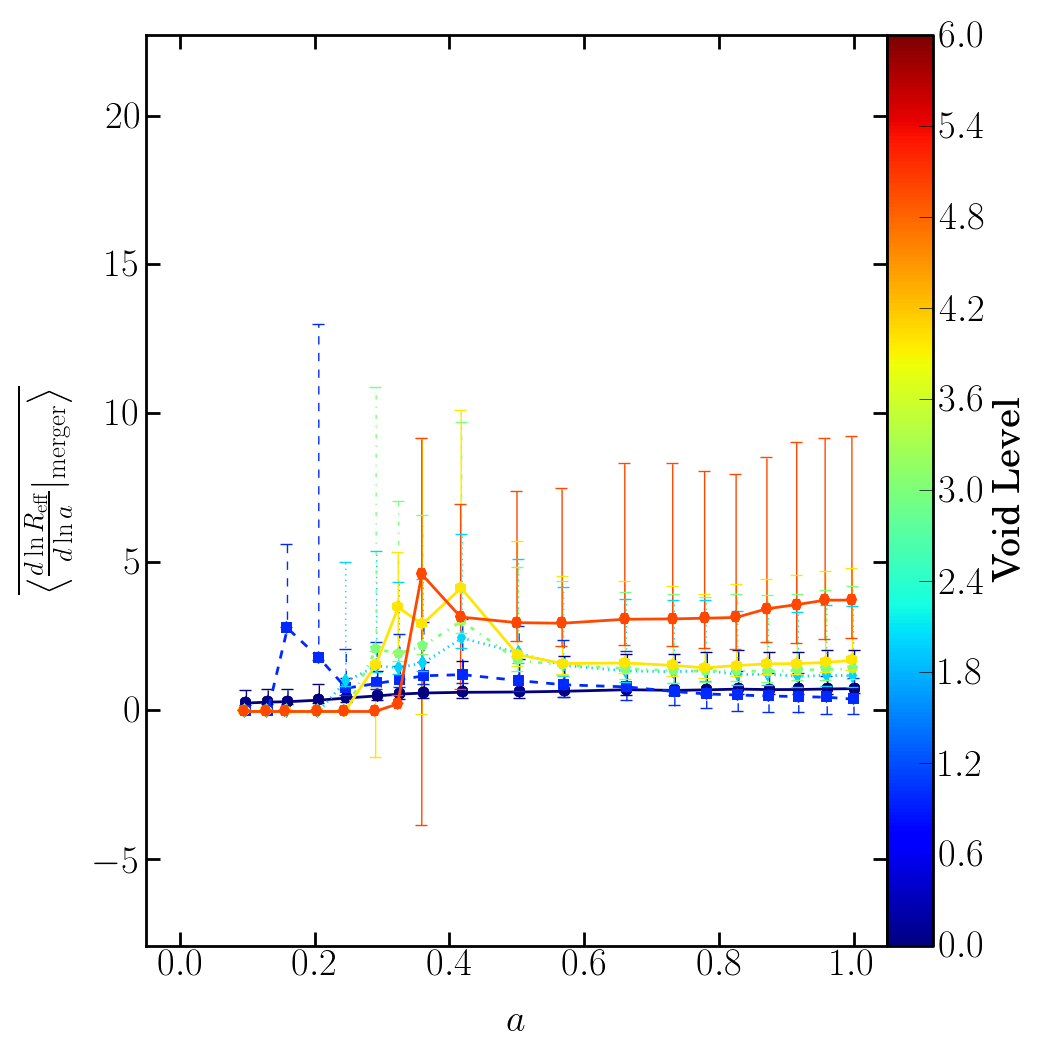}}
  \caption{
           Breakdown of void growth rate into secular (left panel) and
           merging (right panel) components. Shown is the mean growth 
           rate (solid lines) and the maximum spread (dashed error bars)
           for all voids as a function of scale factor. Voids are grouped 
           by their position in the void hierarchy, from the top-most 
           parent voids (blue) to the deepest children void (red). 
          }
\label{fig:growth-breakdown}
\end{figure*}

We break down the growth rates into secular and merging 
components, as we show in Figure~\ref{fig:growth-breakdown}. 
We define the secular growth rate as the growth rate of 
voids which did not experience a merger in that timestep. If instead that 
void did merge with another, its growth from snapshot to snapshot 
is calculated in the average merger growth rate.
Here we plot the mean growth rate over all voids as a function 
of scale factor. We also separate voids into their level in the 
hierarchy so that we may examine the nature of larger parent 
voids and their subvoids separately.

First we notice that the merger growth rate far outweighs the 
secular growth rate by an order of magnitude; voids gain volume typically 
not by growth of the underlying volume but by merging (when it does 
occur) with adjacent voids. The fractional merger growth rate is much larger 
for subvoids deep in the hierarchy than it is for voids higher 
in the tree. Thus, even though a small fraction of 
larger voids experience occasional 
large jumps in their volume, averaged over the entire cohort of voids 
in that tree level 
it is entirely insubstantial. We see for voids deep in the 
hierarchy (that is, subvoids) a peak in the merger growth 
rate at scale factor $0.3$, 
which is the epoch with the highest formation rate where
voids experience a rapid 
restructuring as the hierarchy forms.

Comparatively, the secular growth rate is very small and contributes 
little to the overall growth of voids. Here we see the opposite trend 
as for the merger-based growth rate: the voids highest in the tree 
hierarchy have the 
highest rates, since they are not surrounded by overdense shells that would 
restrict their growth. Again we see a collapsing void 
population.
Indeed, they have been collapsing since 
$a=0.3$, when the top-level voids first formed.
 
\begin{figure} 
  \centering 
  {\includegraphics[type=png,ext=.png,read=.png,width=0.48\textwidth]{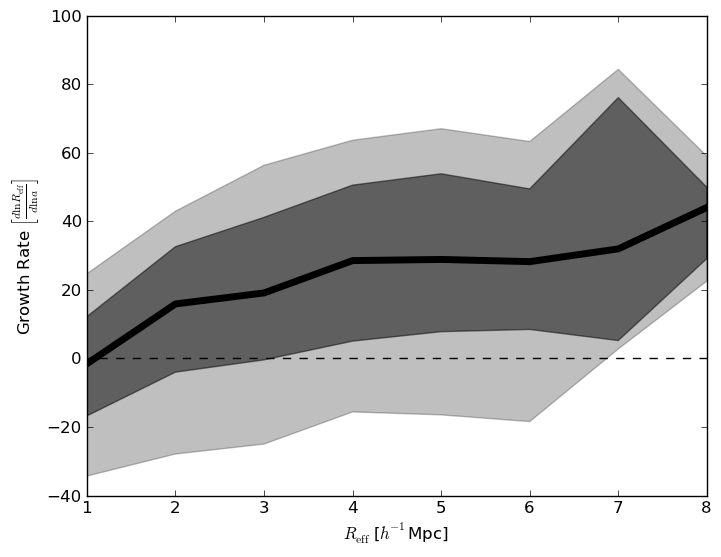}}
  \caption{
            Instantaneous secular (i.e., not merging) 
            growth rate at $a=1.0$ as a function of 
            void effective radius. The solid black line is the median
            growth rate in bins of 1~\hmpc, while the dark and light 
            bands are the 68\% and 95\% binned percentiles in the 
            distributions.
          }
\label{fig:growthrate}
\end{figure}

Finally, Figure~\ref{fig:growthrate} shows the 
instantaneous growth rate at $a=1.0$. This is an interesting statistic
since it allows us to separate collapsing from expanding voids 
based purely on their recent history. Similar analyses have been done 
using the void-matter 
cross-correlation~\citep{Hamaus2013} and the identification of
universal density and velocity profiles~\citep{Hamaus2014,Sutter2013a}. 
While we do not have sufficient 
volume to reliable apply those measures here, we do see a transition 
to overall-collapsing voids around 1-2~\hmpc, although there are subsets of 
collapsing voids at all but the largest scales. 
A cleaner separation is based on 
the position in the void hierarchy, as discussed above.
The clusterings statistics of~\citet{Hamaus2013} estimate a 
compensation scale roughly around 1~\hmpc,
 but this is very uncertain due to the small 
box size. Despite this, we see very good correspondence between 
that result and the point at which the voids, on average, are 
collapsing.

\section{Void Movement}
\label{sec:movement}

\begin{figure} 
  \centering 
  {\includegraphics[type=png,ext=.png,read=.png,width=\columnwidth]{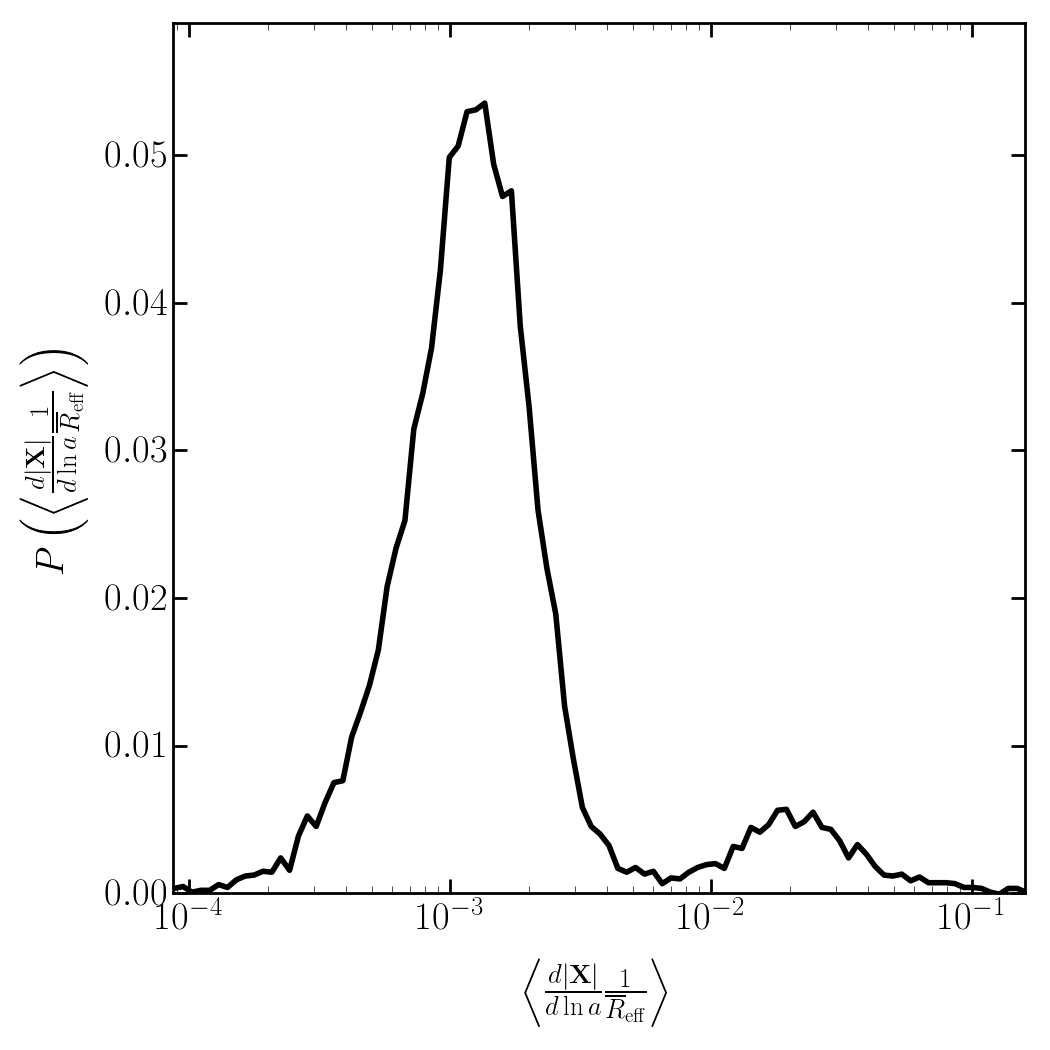}}
  \caption{
            Distribution of mean velocities of the void macrocenter over the 
            void lifetimes.
          }
\label{fig:cmevo}
\end{figure}

Despite the occasional violent merger, all voids experience very 
little movement of their macrocenters (Eq.~\ref{eq:macrocenter}).
Figure~\ref{fig:cmevo} shows the mean void macrocenter velocity, 
which we define as $\left< d |{\bf X}|/d \ln a\right>$, and which 
we express as fractions of the void effective radius at the 
current epoch. The average is taken over the entire void lifetime 
tracing from its current state along the branch of its main 
progenitors.

The distribution of mean macrocenter velocities has two distinct peaks. 
One, at $10^{-3}$, represents the movement of the majority 
of voids, and is remarkably low. This mean velocity gives rise to an 
average displacement of only a tiny fraction of the void radius. 
This is not surprising: once a deep underdensity forms from the 
initial conditions, it is unlikely to move as the voids expand 
into the surrounding cosmic web. 

The second peak, at $1 \times 10^{-2}$, is the small population of 
voids that experience violent mergers. Most of these voids are small 
subvoids residing deep in the void hierarchy. As before, we see that this is 
only a small fraction, roughly five percent, of all voids.
Even in these cases, the mean displacement is very small, indicating
that even when voids experience mergers they are relatively 
gentle, since the macrocenter remains relatively stable.

\section{Void Destruction}
\label{sec:destruction}

\begin{figure} 
  \centering 
  {\includegraphics[type=png,ext=.png,read=.png,width=0.48\textwidth]{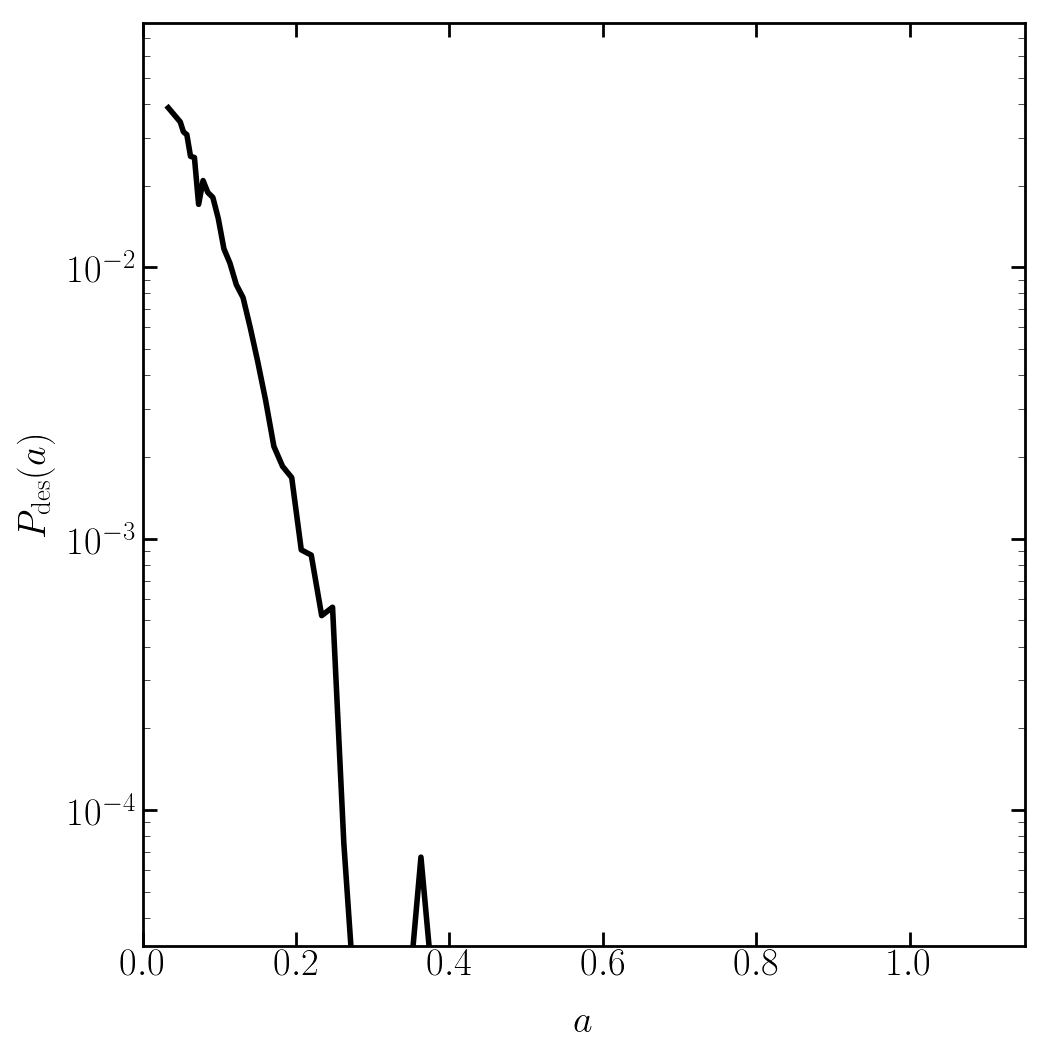}}
  \caption{
           Fraction of the existing void population destroyed as a function of 
           scale factor.
          }
\label{fig:destrate}
\end{figure}

We show the destruction rate, 
or the fraction of the existing void population lost 
in each snapshot
in Figure~\ref{fig:destrate}.
While initially 
high prior to $a=0.3$, 
afterwards the destruction rate drops to essentially zero. 
The initial relatively high destruction rate is not surprising,
since at these early times the basins are just beginning to 
become deep enough to be identified as voids, and there is significant 
noise in the classification of these objects.
However, even at these low scale factors no more than $\sim 4\%$ of voids 
are lost in every snapshot.
But at the same time the formation rate spikes, $a=0.3$, the 
destruction rate plummets. After this epoch, voids that have already 
existed or will eventually form never cease to exist. 
Thus even voids that are collapsing are not squeezed entirely; 
instead they merely become subvoids of larger parent voids.
In fact, the squeezing may help bolster their ability to be 
detected: as the overdense shells around them grow higher, 
the density contrasts increase, allowing the void finder to 
continually detect them. 

\section{Conclusions}
\label{sec:conclusions}
  
We have performed a comprehensive analysis of the life cycle ---
covering formation, mergers, growth, movement, and destruction --- 
of cosmic voids. We have adapted merger tree codes originally designed
to track the evolution of halos to account for the large spatial
extents of voids. By applying this technique to a high-resolution 
$N$-body simulation, we have gained a clear picture of voids as
dynamic objects in the cosmic web. Through the use of a watershed 
void finder, we are able to classify voids according to their position
in a hierarchy and use that to identify key epochs and scales 
in their evolution.

The past life of a cosmic void depends intimately
on its place in the void hierarchy. Voids near the top of the 
hierarchy primarily form at a scale factor of $0.3$, when the
density contrasts in the cosmic web become high enough 
to support their identification and the introduction of dark 
energy shuts off continued structure formation. These higher-level 
voids suffer only minor mergers and tend to maintain consistent growth 
rates over cosmic time. 
In contrast, voids lying deep in the hierarchy continue to form 
and have a somewhat more violent life due to the lower-density 
 nature of their surroundings, but even 
most of these voids have only a single line of 
descent. The location of a void in the hierarchy is more important 
than its size: two voids of equal volume can have radically 
different merger histories depending on their amount 
of substructure.

Voids typically grow at slow rates. However, there is a population
of small collapsing voids.
These voids tend to 
live in overdense environments near filaments and walls, as 
initially pointed out by~\citet{vdW2004}. 
Their overdense surroundings slowly squeeze them as
adjacent larger voids expand. 
This picture is consistent with the theory developed
by~\citet{Sheth2004}, the velocity inflow-outflow analysis 
of~\citet{Ceccarelli2013},
the clustering study of~\citet{Hamaus2013}, and the 
density profile studies of~\citet{Hamaus2014} 
and~\citet{Sutter2013a}.
Despite being slowly crushed, after $a=0.3$, these
voids never get completely destroyed. Instead, they continue to survive
as identifiable voids to the present day.
Thus the void destruction rate does not play a significant role in the 
late-time evolution of voids, and can be ignored 
in theoretical treatments. Additionally, as pointed out 
by~\citet{Russell2013}, the collapsing process is completely negligible 
for all but the smallest voids, although this result is in conflict 
with analyses based on the adhesion approximation~\citep{Sahni1994}.

Finally, voids do not move much throughout their lifetimes. Only small voids in the frothy depths of the hierarchy that undergo several mergers appear to 
have perturbed macrocenters. Even for these most active of voids, they typically only move a few percent of their effective radii.

The combination of small box volume and high resolution limits 
our study to relatively small ($\sim 1-15$~\hmpc) voids, while voids 
in larger simulations and galaxy surveys are typically much larger.
However, recently~\citet{Sutter2013a} were able to show that many void
properties scale as a function of sampling density and galaxy bias.
Thus, properties and characteristics of voids studied in one population 
of tracers can, in principle, be immediately translated to voids 
in another population. Thus the conclusions that we reach in this work 
are generally applicable to voids discovered in other simulations 
and galaxies.

We have examined the properties of voids defined using a watershed 
technique. There are, of course, other plausible definitions of 
voids (see, for example, the comparison work of~\citealt{Colberg2008}).
These different algorithms might give different pictures of 
void histories, especially formation times, since they usually 
impose density thresholds. 
Additionally, there are other approaches to defining merger trees.
We have noticed that volume correlations based on particles can give some 
non-intuitive results: voids that appear to occupy similar positions
(based on their macrocenters and effective radii) may not necessarily 
share any particles. The relationships between particle correlation 
and macrocenter definition should be investigated further. 
However, we have applied other merger tree algorithms, 
such as \textsc{MergerTree} and \textsc{JMerge} (both described 
in~\citealt{Srisawat2013}), and found qualitatively similar 
results.

Overall, voids live far quieter lives than their
overdense counterparts, the halos. Whereas up to $20\%$ of halos
have suffered a recent major merger, voids experience essentially
\emph{no} major mergers throughout their lifetime.
Likewise, while subhalos can be stripped of their mass as they pass through
a larger parent halo, subvoids continue to be identifiable
even when a supervoid forms around them.
The implication is that voids are a much more pure cosmological
probe; the fundamental cosmological signal imprinted from
initial conditions and modified by dark energy is not
corrupted by significant dynamics.
Thus lower-redshift cosmological probes, such as the Alcock-Paczynski
test and void-galaxy cross-correlations, 
will not be affected by recent spurious mergers
in the void population.

\section*{Acknowledgments}

PMS acknowledges
support from NSF Grant NSF AST 09-08693 ARRA, and 
would like to thank Mark Neyrinck, Alice Pisani,
Ravi Sheth, Ben Wandelt, and David Weinberg for 
useful comments and discussions.

PJE would like to thank Krzysztof Bolejko for many useful discussions over tea. PJE is supported by the SSimPL programme and the Sydney Institute for Astronomy (SIfA), DP130100117.

BF is supported by the UK Science and Technology Facilities Council grants ST/K00090/1 and ST/L005573/1.

AK is supported by the {\it Ministerio de Econom\'ia y Competitividad} (MINECO) in Spain through grant AYA2012-31101 as well as the Consolider-Ingenio 2010 Programme of the {\it Spanish Ministerio de Ciencia e Innovaci\'on} (MICINN) under grant MultiDark CSD2009-00064. He also acknowledges support from the {\it Australian Research Council} (ARC) grants DP130100117 and DP140100198. He further thanks Momus for the myth of the bishonen.

All the authors are thankful for the generosity and hospitality of the 
organizers of the Sussing Merger Trees 
Workshop~\footnote{{http://popia.ft.uam.es/SussingMergerTrees}}, held in July 2013, where this project started, and 
especially Frazer Pearce and Peter Thomas, who provided valuable 
comments.
The Sussing Merger Trees Workshop was supported by the European 
Commission's Framework Programme 7 through the Marie Curie Initial Training Network CosmoComp (PITN-GA-2009-238356). This also provided fellowship support for AS.

The authors contributed in the following ways to this paper: 
AK, CS, and AS were members of the SOC that organized the workshop from which this project originated. They helped design the comparison, plan, and organize the data. 
The analysis presented here was performed by PE, PMS, BF, JO, and NH 
and the paper was written by PMS. 
The other authors contributed towards the content of the paper and helped to proof-read it.

\footnotesize{
  \bibliographystyle{mn2e}
  \bibliography{voidtrees}		
}

\end{document}